\documentclass[final,3p,times,twocolumn]{elsarticle}

\usepackage{amssymb}
\usepackage{color}
\usepackage{url}
\usepackage{nameref}
\usepackage{colortbl}
\usepackage{newfloat}
\usepackage[most]{tcolorbox}
\biboptions{sort&compress}
\newcommand{\makecell}[2][l]{\begin{tabular}{@{}#1@{}}#2\end{tabular}}

\usepackage{paralist}

\definecolor{Gray}{gray}{0.95}

\DeclareFloatingEnvironment[
  name = Box
]{mybox}

\journal{Current Opinion in Structural Biology}

\begin{document}

\begin{frontmatter}

\title{Integrating experimental data with molecular simulations to investigate RNA structural dynamics}

\author[add1,add2]{Mattia Bernetti}
\ead{Mattia.Bernetti@iit.it}

\author[add3]{Giovanni Bussi\corref{cor1}}
\ead{bussi@sissa.it}

\address[add1]{Computational and Chemical Biology, Italian Institute of Technology, 16152 Genova, Italy}
\address[add2]{Department of Pharmacy and Biotechnology, University of Bologna, 40126 Bologna, Italy}
\address[add3]{Scuola Internazionale Superiore di Studi Avanzati, via Bonomea 265, 34136, Trieste, Italy}

\cortext[cor1]{Corresponding author}

\begin{abstract}
Conformational dynamics is crucial for ribonucleic acid (RNA) function.
Techniques such as nuclear magnetic resonance, cryo-electron microscopy, small- and wide-angle X-ray scattering,
chemical probing, single-molecule F\"orster resonance energy transfer or even
thermal or mechanical denaturation experiments probe RNA dynamics at different time and space resolutions.
Their combination with accurate atomistic molecular dynamics (MD) simulations
paves the way for quantitative and detailed studies of RNA dynamics.
First, experiments provide a quantitative validation tool for MD simulations.
Second, available data can be used to refine simulated structural ensembles to match experiments.
Finally, comparison with experiments allows for improving MD force fields that are transferable to new systems for which
data is not available.
Here we review the recent
literature and provide our perspective on this field.
\end{abstract}


\end{frontmatter}

\copyright 2022. This manuscript version is made available under the CC-BY-NC-ND 4.0 license https://creativecommons.org/licenses/by-nc-nd/4.0/.
This is an accepted manuscript (postprint). The published journal article is available at \url{https://doi.org/10.1016/j.sbi.2022.102503}.

\sloppy

\section*{Introduction}
\label{sec:introdution}

\begin{figure}
\begin{center}
\includegraphics[width=0.45\textwidth]{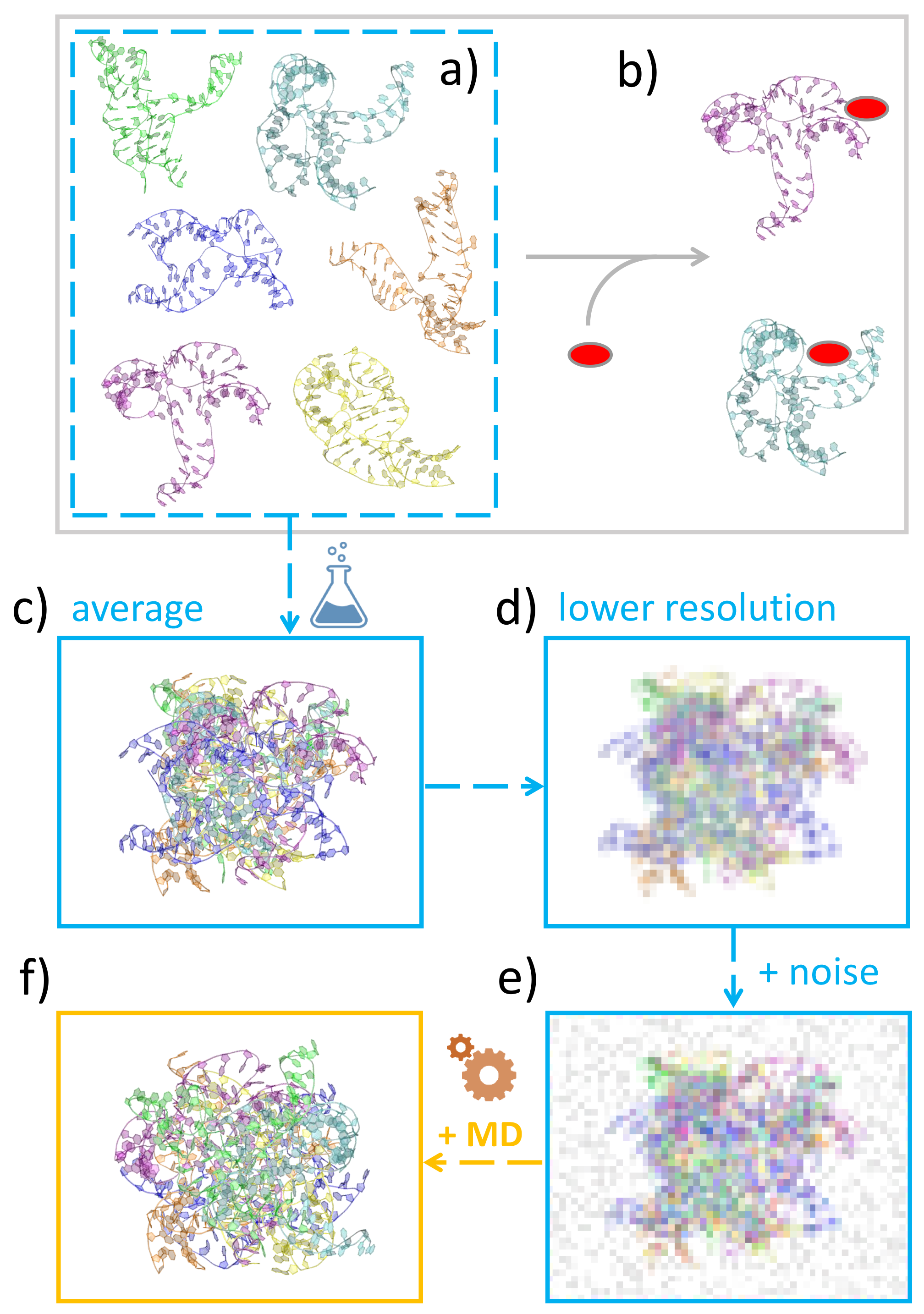}
\end{center}
\caption{
\label{fig:rna-dynamics}
RNA dynamics. RNA molecules in solution sample multiple conformations (a). Upon binding with 
a small ligand or another macromolecule (in red), a subset of these conformations is selected
(conformational selection) or a few additional ones are induced (induced fit) (b).
Being able to characterize the unbound and bound structural ensembles is central in studying interactions.
Solution experiments usually report information that is averaged over the conformational
ensemble (c) and only contains a lower resolution description (d), usually affected by noise (e).
Accurate modeling tools are required to deconvolve dynamics and reconstruct the original ensembles (f).
Representative RNA structures were adapted from those generated in Ref.~\cite{bernetti2021reweighting}.
}
\end{figure}

Ribonucleic acid (RNA) structure is often necessary for function \cite{vicens2022thoughts}.
However, RNA structures must not be considered as static pictures.
RNA dynamics, i.e., the capability to explore multiple conformations,
is indeed crucial as it is 
required for (and modulated by) binding
with small molecules, either metabolites \cite{cheng2022cotranscriptional}
or drug-like molecules \cite{ganser2018high}, or macromolecules \cite{abeysirigunawardena2017evolution}
(see Figure~\ref{fig:rna-dynamics}a--b).
Many experimental techniques have been developed to probe RNA structure.
X-ray crystallography \cite{westhof2015twenty} provides very detailed pictures but restricts dynamics
through crystal packing effects, thus with limited applicability when dealing with flexible RNAs.
Conversely, solution methods such as nuclear magnetic resonance (NMR) \cite{liu2021developments} and cryo-electron microscopy \cite{kappel2020accelerated} allow for probing RNA dynamics,
at the price of a lower atomic resolution.
Additionally, coarser-grained experimental techniques, such as
small- (or wide-) angle X-ray scattering (SAXS/WAXS) \cite{chen2016saxs}, chemical probing \cite{weeks2021shape},
single-molecule F\"orster resonance energy transfer \cite{steffen2021fretraj} or,
to some extent, thermal \cite{andronescu2014determination} or mechanical \cite{rissone2022stem} denaturation experiments
are also capable to access RNA dynamics. A common trait of these techniques
is that they report possibly noisy ensemble averages over the set of accessible structures
(see Figure~\ref{fig:rna-dynamics}c--e).
The dynamical information is thus present but must be deconvolved.
Molecular dynamics (MD) simulations \cite{sponer2018rna} enable RNA dynamics to be predicted and characterized at the atomistic
scale.
Limitations on the accessible timescales can be alleviated by taking advantage of
enhanced sampling methods \cite{henin2022enhanced}.
However, despite recent improvements, the accuracy of the models used to describe interactions (force fields) for RNA is still limited,
in particular for some structural motifs \cite{mrazikova2020uucg}.
Nevertheless, MD simulations can be fruitfully applied
also when current force fields are not predictive, by suitable integration of experimental data (see Figure~\ref{fig:rna-dynamics}f).

In this review, we discuss how MD simulations and experimental data can be combined to investigate
the conformational dynamics of RNA molecules, covering recent studies where
simulations are either quantitatively
validated on solution experiments or integrated with experimental information.
We will not discuss structure prediction methods, whose typical goal is to find a single
relevant structure.  The survey will be limited to the past few years.

\section*{Combining molecular simulations and solution experiments}
\label{sec:combining}

\subsection*{Integration strategies}
\begin{figure*}
\begin{center}
\includegraphics[width=\textwidth]{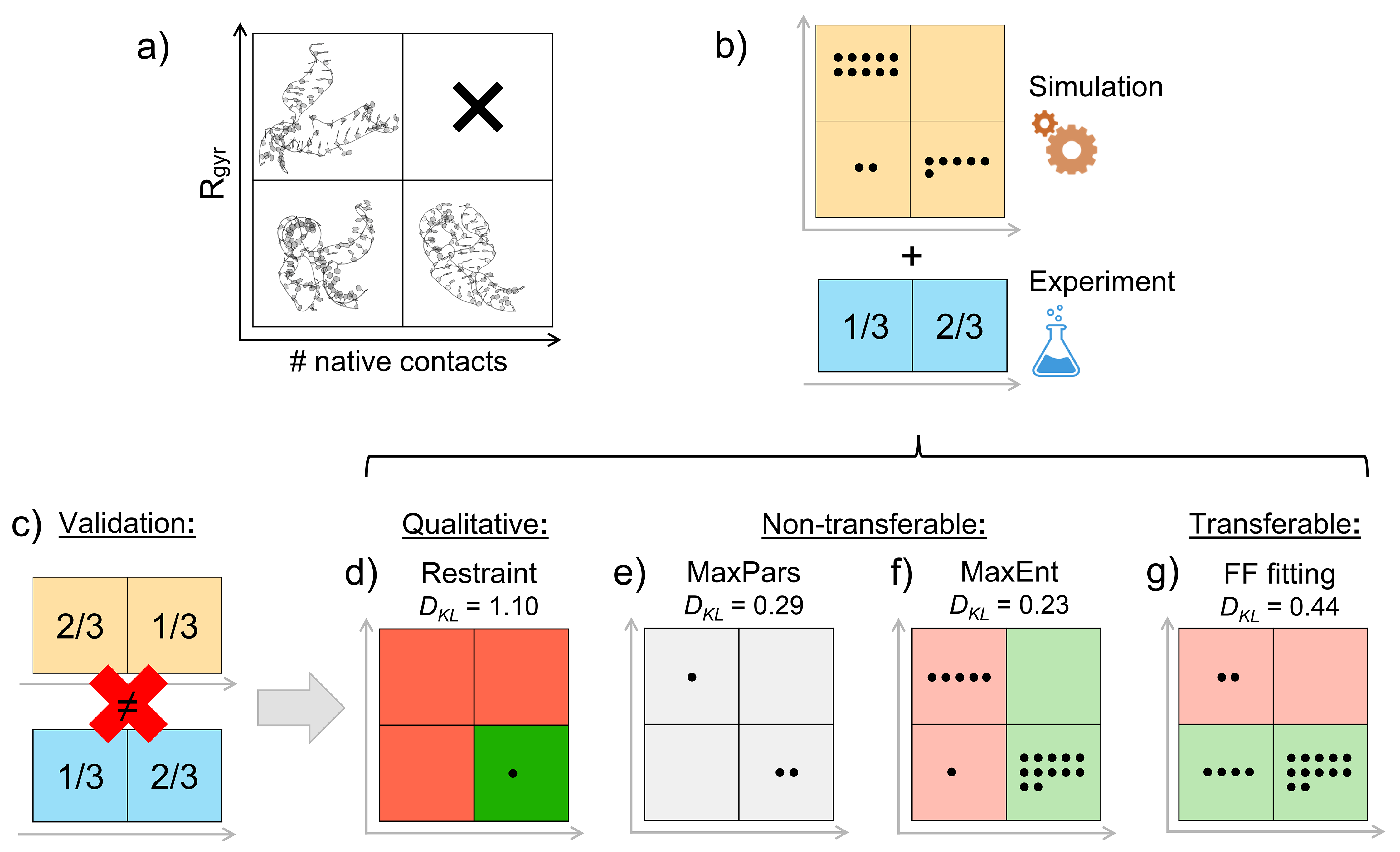}
\end{center}
\caption{
\label{fig:combining-md-exp}
Integrating simulations and experiments.
We consider a molecule with three available conformations (a):
extended (E, large gyrations radius, $R_{gyr}$, and no native contacts formed),
compact (C, small $R_{gyr}$ and no native contacts formed), and native (N, small $R_{gyr}$ and many native contacts formed, 
highlighted by the green shade and box contour).
Native contacts are assumed to be those present in a reference structure.
An MD simulation results in populations of the three states as reported in the yellow square,
proportional to the number of dots,
and the experiment suggests that the population of the native state is 2/3 (b).
The comparison of the MD simulation with the experiment indicates that the MD ensemble is incorrect (c).
One could qualitatively restrain the simulation to native (d).
More quantitatively, one can choose a minimal number of conformations that agree with the experiment (maximum parsimony, MaxPars, e),
or minimally modify the MD ensemble (maximum entropy, MaxEnt, f).
Both these approaches are not transferable in that they 
only allow extracting information about the system for which experiments were performed.
Finally, a force-field (FF) fitting strategy could be used to obtain transferable corrections (g).
In this example, we assumed that the water model or the torsional parameters are modified, impacting the relative population
of the extended state.
Green and red shades indicate which states are favored and disfavored by the experimental data, respectively.
Kullback-Leibler divergences ($D_{KL}$) from the original MD ensemble, computed using the populations of the three states,
are reported for the four generated ensembles.
Representative RNA structures were adapted from those generated in Ref.~\cite{bernetti2021reweighting}.
}
\end{figure*}

Multiple strategies can be used to integrate MD and experimental information
(see \cite{orioli2020learn} for a recent review and Figure~\ref{fig:combining-md-exp} for a schematic).

First,
experiments can be used to quantitatively \emph{validate} MD-generated ensembles (Figure~\ref{fig:combining-md-exp}c).
      For instance, testing multiple force fields, it is possible to determine which is the most trustable one, thus generating
      information transferable to other systems.

Second,
experiments can be used to \emph{improve generated ensembles} using either:
      \begin{compactitem}
      \item Qualitative methods, where data are used to generate initial structural models or to restrain the simulations without
      explicitly modeling the experimental dynamics (Figure~\ref{fig:combining-md-exp}d).
      \item Quantitative methods, ranging from maximum parsimony, such as sample-and-select (Figure~\ref{fig:combining-md-exp}e),
           to maximum entropy (Figure~\ref{fig:combining-md-exp}f); see Ref.~\cite{medeiros2021comparison} for an overview.
      Results are not transferable to other systems but can be used to (a) translate the experiment to detailed structural information and (b)
      predict new experiments on the same system.
      Methods can be used both for reweighting/analyzing existing trajectories or to
      restrain them on-the-fly, the latter approach being more efficient when the original simulated
      ensembles are farther from the experiment.
      \end{compactitem}

Finally,
experiments can be used to \emph{improve force fields} \cite{frohlking2020toward} (Figure~\ref{fig:combining-md-exp}g).
      Force-field parameters can be first trained using reweighting techniques that enable predicting the impact of parameter modifications,
      and then validated with new simulations.
      Results are expected to be transferable to other systems for which the experiments are not available yet.
      To enhance transferability, one should train the force field on a large and heterogeneous set of systems.

\subsection*{Critical issues}
Several critical issues should be taken into account when comparing simulations and experiments.
First, the magnitude of the experimental error should be considered. An agreement between simulation
and experiment that is better than the uncertainty of the experiment itself should be seen as
an indication of overfitting.
Second, the \emph{forward models}, i.e. the equations to back-calculate the experiment from the MD-simulated structures, are often empirically parametrized
and thus subject to systematic errors.
Then, back-calculated experiments are affected by statistical errors due to the finite length of the simulations.
Comparing experiments with non-converged simulations might lead to
incorrect assessments. When multiple structures are required to
explain experimental results, visiting
them in an accessible simulation time almost invariably requires enhanced sampling techniques.
Finally, no experimental or computational technique is fool-proof,
and experience is required to properly assess the relative merit of different approaches.

\section*{Recent applications to RNA dynamics}
\label{sec:applications}

Here we review recent applications where MD and experiments were combined to study RNA dynamics.
All the discussed applications are summarized in Table~\ref{table}.

\subsection*{Using experimental data for validation or force-field selection}

In several papers, quantitative comparisons with reference experimental data have been used to assess different force fields.
Analyses were based on NMR data for
dinucleoside monophosphates \cite{taghavi2022evaluating}, tetranucleotides \cite{zhao2020nuclear}, or hexanucleotides \cite{zhao2020nuclear,zhao2022nuclear}
and involved a couple 
of force fields, typically a widely adopted one, used as a reference, and a newer variant tested.
NMR data on tetramers were used in Ref.~\cite{mlynsky2020fine}
to scan a wide range of parameters for hydrogen-bond fixes.
Two-way junctions, as well as single-stranded RNAs, were used to test corrections to non-bonded
terms where RNA-RNA and RNA-solvent interactions were tuned separately \cite{he2022refining}.
Here, specific sets of parameters were compared to find those better reproducing SAXS data
and then validated against NMR data.
The tests reported in Refs.~\cite{mlynsky2020fine,he2022refining} were done in a spirit that, although based on enumeration rather than on optimization, resembles that of methods
discussed in Section \emph{\nameref{sec:force-field-fitting}}.
In other papers, alchemical calculations have been validated with binding affinities \cite{tanida2020alchemical} or denaturation free energies \cite{hurst2021deciphering},
testing a single set of force-field parameters against experiments.

\subsection*{Integrating experimental data and simulations to improve ensembles}

Two recent works used restraints based on the secondary structure obtained from either NMR \cite{baronti2020base} or chemical probing \cite{cheng2022cotranscriptional} data, employing a two-dimensional replica-exchange method.
Here, replicas in which the experimental information is used to enforce known base pairings are coupled with unbiased replicas.
In this manner, the correct base pairing is encouraged but not enforced, and dynamics is preserved.
Base pairing information obtained from chemical probing and NMR data was also used to construct initial structures in Ref.~\cite{bottaro2021conformational}.
Enhanced sampling simulations were then conducted without restraints,
but the experimental base pairing was preserved by choosing a suitable replica-exchange approach.
Contacts predicted by nuclear Overhauser effect (NOE) data have been qualitatively used as instantaneous restraints when
equilibrating protein-RNA complexes \cite{krepl2021recognition,clery2021structure}, followed by unrestrained MD simulations
with the exception of a few simulations employing soft NOE restraints with a limited impact on the dynamics.
In Ref.~\cite{krepl2021recognition} the authors validated
quantitatively the results against experimental RNA-protein affinities and effects of chemical modifications.
In all these works, for an infinitely long simulation, one should
expect the result to be unaffected by the presence of the experimental information. Hence, these methods should be better seen as sampling or modeling tools
rather than integrative methods.
Experimentally derived secondary structures were used also in Ref.~\cite{yu2021computationally}.
Here, structures obtained from co-transcriptional chemical-probing experiments
were enforced during the simulation, using an \emph{ad hoc}  procedure to avoid the system being stuck in non-productive pathways.
In this case, the experimentally-derived secondary structure was enforced in the MD-generated structures. However, dynamics was not quantitatively controlled.

Cryo-electron-microscopy experiments have been used in so-called flexible-fitting MD simulations
to refine structures \cite{bonilla2021viral,bonilla2022cryo}.
In this setting, MD simulations are used as a modeling tool to generate structures compatible with experiments,
but the quantitative degree of dynamics compatible with the experimental density maps is not
explicitly controlled.

Several works used the maximum entropy principle to enforce ensemble averages quantitatively.
The simulated ensemble of a UUCG tetraloop was reweighted to match a set of NMR data, including
NOEs, exact NOEs, $^3J$ scalar couplings, and solvent paramagnetic relaxation enhancement \cite{bottaro2020integrating}. The MD simulation suggested an alternative
structure for the loop. Including available data suggested a lower but not negligible population of this alternative structure.
Interestingly, different data sets led to different results.
A similar approach was used in Ref.~\cite{bergonzo2022conformational}, where reweighting was conducted using either NMR or SAXS data on ensembles
generated for an RNA hexamer. Agreement with NMR did not necessarily result in agreement with SAXS, and vice versa,
suggesting that multiple, independent experimental observables are important to assess 
the accuracy of heterogeneous structural ensembles.
Multiple force-field parameters for atomistic MD were compared.
NOE data were also used in Ref.~\cite{reisser2020conformational} to construct heterogeneous ensembles for an RNA hairpin
found in a non-coding RNA.
In this case, reweighting was not sufficient and experimental data were thus enforced during the simulation.
A maximum parsimony approach was then employed to generate ensembles compatible with the experiment
but composed of structures belonging to a limited number of
structural clusters. This approach has the advantage that it can be used to obtain
ensembles that are simpler to describe, at the price of decreasing dynamics and modifying more the ensemble relative to the one simulated with MD.
SAXS data were used in Ref.~\cite{bernetti2021reweighting} to quantify the population of compact and extended conformations
in a larger structured RNA. Here, different forward models for SAXS were tested, showing that
both solvent and dynamical effects are crucial to match experimental data. In this case, enhanced sampling
allowed to have both compact and extended structures in the pool of conformations used for reweighting.

Other available methods for ensemble refinement are based on the maximum parsimony principle,
and include for instance the popular sample-and-select and ensemble optimization approaches.
In Ref.~\cite{he2021structural}, WAXS data were used to select conformations from MD simulations of RNA and DNA duplexes.
Simulations performed at a higher temperature were used to enhance sampling in cases where conventional MD at room temperature
was not providing conformations capable of rationalizing the experimental spectrum.
A similar approach was used in Ref.~\cite{chen2022insights} to analyze RNA triplexes.
Here, WAXS-driven MD was used to tackle cases where high-temperature sampling was not sufficient.
WAXS data were thus used both on-the-fly and a posteriori.
The sample-and-select method was also used to analyze MD simulations of TAR RNA to match measured residual dipolar couplings
\cite{shi2020rapid}. Interestingly, in this case, the authors reported that an ensemble generated with a fragment assembly method (FARFAR)
was matching experiments better than atomistic MD. It is unclear if this is due to the higher accuracy of FARFAR or to better sampling.
Furthermore, these results might be contrasted with the opposite finding in two papers discussed above \cite{bottaro2021conformational,bergonzo2022conformational}.

\subsection*{Integrating experimental data and simulations to improve force fields}
\label{sec:force-field-fitting}

The quantitative mismatch between simulation and experiment has been also used to refine force-field parameters.
In Ref.~\cite{frohlking2022automatic}, simulations of a tetramer were compared with
NMR experimental data.
A minimization strategy was then used 
to optimize hydrogen-bond corrections,
additionally including in the cost function the comparison between tetraloop simulations and static information from X-ray crystallography.
The resulting force field was capable to stabilize the difficult UUCG tetraloop and reproduce solution data for
additional tetramers that were not used during training.

A torsional potential and, more notably, partial charges were fine-tuned in Ref.~\cite{piomponi2022molecular} for a modified
nucleobase. Parameters were tuned with an optimization procedure to match $\Delta \Delta G$s
measured in thermal denaturation experiments. Whereas modifying charges leads to long-range effects that might make
the reweighting procedure not efficient, tiny modifications were sufficient to significantly improve the agreement with the experiments.
This work shows that it is possible to significantly improve agreement with experiment by tuning partial charges without
leaving the fixed-charge approximation. However, the explicit modeling of polarization effects
might be necessary to enable a further step forward in the accuracy and transferability of RNA force fields \cite{salsbury2021recent}.

Reference \cite{chen2022rna} used a two-dimensional correction on the $\alpha$ and $\zeta$ backbone torsional angles
to decrease the population of intercalated structures for RNA tetramers.
The procedure was carried out with
reweighting, to virtually test a large number of parameter sets, but without explicitly computing derivatives
of the cost function with respect to the parameters. It is worth noting that a similar effect can be partly obtained using
a completely different approach, namely switching to a different water model.
Indeed, force-field refinement strategies do not lead to unique results and require manual intervention to decide
which parameters should be optimized.

\begin{table*}[]
\caption{Recent applications where MD and experiments were combined to study RNA dynamics.
For each of the discussed articles, the table reports which method was used to integrate MD simulations and experiments,
which type of experiment was performed, which is the corresponding structural information,
and the references to the simulation work discussed here and to the primary source for experimental data.
Abbreviations:
maximum entropy (MaxEnt),
maximum parsimony (MaxPars),
nuclear magnetic resonance (NMR),
nuclear Overhauser effect (NOE),
exact nuclear Overhauser effect (eNOE),
residual dipolar coupling (RDC),
solvent paramagnetic relaxation enhancement (sPRE),
small-angle X-ray scattering (SAXS),
wide-angle X-ray scattering (WAXS),
cryo-electron microscopy (cryo-EM).
\label{table}
}
\begin{center}
\begin{small}
\begin{tabular}{lllcc}
\hline
\\[-1em]
Integration method & Data                   & Information provided                                       & Simulation & Experiment                                                          \\
\\[-1em]
\hline
\\[-1em]
\multicolumn{5}{c}{\emph{Validation against experiment or force field selection}} \\
\\[-1em]
\hline
\rowcolor{Gray}
Validation         & $^3J$ couplings           & Sugar conformation                                                         & \cite{taghavi2022evaluating} & \cite{ezra1977conformational}                         \\
Validation         & $^3J$ couplings and NOEs           & \makecell[l]{Sugar and backbone conformation,\\ contacts}                  & \cite{zhao2020nuclear}       & \cite{zhao2020nuclear,condon2015stacking}                             \\
\rowcolor{Gray}
Validation         & $^3J$ couplings and NOEs           & \makecell[l]{Sugar and backbone conformation,\\ contacts}                  & \cite{zhao2022nuclear}       & \cite{zhao2022nuclear}                                                           \\
Validation         & $^3J$ couplings and NOEs           & \makecell[l]{Sugar and backbone conformation,\\ contacts}                  & \cite{mlynsky2020fine}       & \cite{yildirim2011benchmarking,tubbs2013nuclear,condon2015stacking} \\
\rowcolor{Gray}
Validation         & $^3J$ couplings and SAXS           & \makecell[l]{Sugar and backbone conformation,\\ inter-helical orientation} & \cite{he2022refining}        & \cite{condon2015stacking,chen2019conformations}                     \\
Validation         & Equilibrium filtration & Binding affinities                                         & \cite{tanida2020alchemical}  & \cite{jenison1994high} \\
\rowcolor{Gray}
Validation         & Thermal denaturation           & Mutational free energies                                   & \cite{hurst2021deciphering}  & \cite{roost2015structure} \\
\hline
\\[-1em]
\multicolumn{5}{c}{\emph{Qualitative integration of experimental dynamics}} \\
\\[-1em]
\hline
\rowcolor{Gray}
\makecell[l]{Secondary structure \\2D replica exchange} & NMR                      & Secondary structure &\cite{baronti2020base}            & \cite{baronti2020base}                  \\
\makecell[l]{Secondary structure \\2D replica exchange} & Chemical probing         & Secondary structure &\cite{cheng2022cotranscriptional} & \cite{cheng2022cotranscriptional}                  \\
\rowcolor{Gray}
\makecell[l]{Restraints \\ during equilibration}                     & NMR and chemical probing & Secondary structure &\cite{bottaro2021conformational}  & \cite{wacker2020secondary} \\
\makecell[l]{Restraints \\ during equilibration}                     & NOEs                     & Contacts            &\cite{krepl2021recognition}       & \cite{krepl2021recognition}                  \\
\rowcolor{Gray}
\makecell[l]{Restraints \\ during equilibration}                     & NOEs                     & Contacts            &\cite{clery2021structure}         & \cite{clery2021structure}                  \\
Restraints                                                           & Chemical probing         & Secondary structure &\cite{yu2021computationally}      & \cite{yu2021computationally}                  \\
\rowcolor{Gray}
Flexible fitting                                                     & Cryo-EM                  & Density map         &\cite{bonilla2021viral}           & \cite{bonilla2021viral}                 \\
Flexible fitting                                                     & Cryo-EM                  & Density map         &\cite{bonilla2022cryo}            & \cite{bonilla2022cryo}                  \\
\hline
\\[-1em]
\multicolumn{5}{c}{\emph{Quantitative integration of experimental dynamics (non-transferable ensemble refinement)}} \\
\\[-1em]
\hline
\rowcolor{Gray}
MaxEnt reweighting & \makecell[l]{$^3J$ coupling, NOEs, eNOEs,\\RDCs, sPRE} & \makecell[l]{Sugar and backbone conformation, \\contacts, dipole orientations,\\solvent exposure}
              & \cite{bottaro2020integrating}     & \cite{nozinovic2010high,borkar2017simultaneous,hartlmueller2017rna,nichols2018high} \\
MaxEnt reweighting & \makecell[l]{Chemical shifts, $^3J$ coupling,\\NOEs, SAXS} & \makecell[l]{Sugar and backbone conformation, \\contacts, structure compactness}
              & \cite{bergonzo2022conformational} & \cite{bergonzo2022conformational,zhao2020nuclear}  \\
\rowcolor{Gray}
\makecell[l]{MaxEnt and MaxPars \\ on-the-fly + reweighting} & NOEs               & Contacts &\cite{reisser2020conformational}  & \cite{podbevvsek2018structural} \\
MaxEnt reweighting              & SAXS & Structure compactness &\cite{bernetti2021reweighting}    & \cite{welty2018divalent} \\
\rowcolor{Gray}
Sample-and-select               & WAXS & Pairwise distance correlations &\cite{he2021structural}           &  \cite{he2021structural}\\
\makecell[l]{Sample-and-select \\ + WAXS driven} & WAXS & Pairwise distance correlations & \cite{chen2022insights}             &\cite{chen2022insights}\\
\rowcolor{Gray}
Sample-and-select               & RDCs & Dipole orientations &\cite{shi2020rapid}               & \cite{zhang2007visualizing,salmon2013general} \\
\hline
\\[-1em]
\multicolumn{5}{c}{\emph{Quantitative integration of experimental dynamics (transferable force-field refinement)}} \\
\\[-1em]
\hline
\rowcolor{Gray}
\makecell[l]{Force-field refinement\\(hydrogen bonds)} & $^3J$ coupling, NOEs & \makecell[l]{Sugar and backbone conformation,\\contacts} & \cite{frohlking2022automatic}     &\cite{yildirim2011benchmarking,condon2015stacking} \\
\makecell[l]{Force-field refinement\\(charges and torsion)} & Thermal denaturation & Mutational free energies & \cite{piomponi2022molecular}      &\cite{roost2015structure,kierzek2022secondary} \\
\rowcolor{Gray}
\makecell[l]{Force-field refinement\\(torsions)} & NMR & Population of intercalated structures &  \cite{chen2022rna}                &\cite{condon2015stacking} \\
\hline
\end{tabular}
\end{small}
\end{center}
\end{table*}

\section*{Discussion and perspective}
\label{sec:discussions}

The integration of experimental data and atomistic MD simulations is a growing field of research.
Herein, we gave an overview of the recent works done on this subject in the context of RNA molecules, summarized in Table~\ref{table}.
A significant fraction of the papers discussed here were based on experimental data
previously made available by other authors, also listed in Table~\ref{table}.
In this respect, the effort to measure and publish data for
systems that sometimes are not even biologically relevant is of great value
for groups developing and applying integrative methods.
Moreover, we stress that the interpretation of computational and experimental work is not easy, and
that collaboration between computational and experimental groups is often required.
In the following, we outline our perspective for this field, summarized in Box~1.

An important limitation of the current approaches based on explicit solvent MD simulations is their significant computational cost.
Enhanced sampling methods are thus increasingly adopted in the community.
However, current applications are either addressing unstructured oligomers or larger but relatively structured RNAs.
The simulation of highly flexible long non-coding RNAs might result in further challenges and require new methodological improvements.
The limited accuracy of existing force fields
makes it necessary to include many data points to construct realistic ensembles.
Experimental groups have been traditionally employing implicit-solvent modeling for structure determination
and optionally short explicit-solvent refinements which can only account for minor conformational dynamics
(see, e.g., Refs.~\cite{nozinovic2010high,roost2015structure,hartlmueller2017rna,nichols2018high,podbevvsek2018structural},
for NMR, and Refs.~\cite{bonilla2021viral,bonilla2022cryo}, for cryo-electron microscopy).
Atomistic force fields
are likely more accurate,
and a transition to using extensive explicit-solvent MD to interpret their data is already undergoing in several experimental laboratories.

Whereas most of the applications to date are related to non-transferable methods, where MD is used to
deconvolve available experimental data, a few groups are trying to use experimental data to systematically
improve force fields in a transferable manner. In this sense, force fields can be seen as  energy-based
machine-learning models \cite{du2019implicit}. Training should be done with examples that are representative
of what one wants to reproduce. If the aim is to reproduce and predict experimental
observables, experimental data should be used during training. The physics-based architecture
of the force-field model, though limited by the lack of important aspects such as multi-body and polarization effects,
can improve model transferability. However, to ensure the correctness of the microscopic models
and to obtain force fields that properly describe conformations not yet seen in an experiment, a suitable
combination of solution data, crystallographic information, and bottom-up data from quantum chemistry is required.
Differences in experimental settings, such as ionic strength, temperature, and crystal packing,
should be properly taken into account by performing reference simulations in appropriate conditions or even in
crystalline environments \cite{cerutti2019molecular}.
The ambitious Open Force Field initiative is trying to use these ideas on
large databases of heterogeneous systems, but to date this strategy has only been used to parametrize small molecules \cite{qiu2021development}.
Experimental data on macromolecules will be required to push this approach forward to protein and RNA systems.
Importantly, more commonly used non-transferable approaches such as maximum entropy or maximum parsimony integration methods
will benefit from improved force fields used to generate prior ensembles.
Kinetic data might also provide an extremely valuable source of information as they can probe transition states
\cite{han2022rational}. In theory, they would give access to free-energy barriers that are not seen in
equilibrium experiments. However, integrating them in an ensemble or force-field refinement is nontrivial
and is an active field of research (see, e.g., \cite{brotzakis2021method}).

\begin{mybox}
\label{box}
\begin{tcolorbox}[colback=gray!5!white,colframe=gray!75!black]
Box 1:
Our wish list for the next few years.
\begin{compactitem}
\item More systematic use of explicit solvent and enhanced sampling simulations
in combination with solution experiments.
\item Further improvements in RNA force fields trained to reproduce solution experiments.
\item Integration of kinetic data in ensemble or force-field refinement.
\item Systematic optimization of forward models to back-calculate experimental observables.
\item Development of new forward models to link chemical-probing experiments and three-dimensional dynamics.
\item Explicit reconstruction of RNA dynamics from cryo-electron microscopy data.
\item Simultaneous integration of experiments that probe RNA dynamics at different levels.
\item Combination of experiments and deep-learning structure-prediction methods.
\end{compactitem}
\end{tcolorbox}
\end{mybox}

Another critical issue is the parametrization of forward models used to connect ensembles and experiments.
For example, several sets of Karplus equations have been proposed to back-calculate $^3J$ scalar couplings
(see, e.g., the supporting information of Ref.~\cite{condon2015stacking}), which would lead to different results.
Some of these models were developed 
considering reference static structures and might benefit from a revisiting with dynamics in mind \cite{lindorff2005interpreting}.
Current Karplus equations, with their uncertainty, can thus be used to validate structures but might not
be accurate enough for extracting quantitative populations.
For SAXS experiments, different models include solvent effects to a different degree \cite{bernetti2021comparing}.
Chemical probing experiments are extremely cheap and scalable. Although several attempts to rationalize their signal
in terms of atomistic structure and dynamics have been done in the past years
\cite{pinamonti2015elastic,mlynsky2018molecular,hurst2018quantitative,frezza2019interplay,hurst2021sieving}, quantitative forward models are not available yet so their inclusion in MD simulations is usually done in terms of secondary-structure constraints
\cite{yu2021computationally,bottaro2021conformational,cheng2022cotranscriptional}.
Thermal denaturation experiments are also cheap and scalable, and can provide quantitative reference populations.
Single-molecule F\"orster resonance energy transfer (smFRET) experiments provide access to dynamics.
Notably, there has been some recent improvement in the back-calculation of smFRET from RNA ensembles \cite{steffen2021fretraj},
with an application reported for an RNA tetraloop receptor \cite{erichson2021fret}.
Cryo-electron-microscopy data have been mostly used for RNA structure determination so far. Methods explicitly modeling conformational dynamics
in the back-calculation of density maps
have been used in protein systems \cite{garibsingh2021rational} and could enable a more accurate description of RNA dynamics.

Some of the mentioned experimental techniques can probe dynamics exclusively at a local or at a global scale,
e.g., chemical probing and $^3J$ scalar couplings vs residual-dipolar-couplings, smFRET and SAXS.
Given the ubiquitous risk of overfitting, we suggest that, whenever possible, complementary information should be 
combined so as to enhance the dynamical description at multiple scales, as it has been done for instance
in Refs.~\cite{he2022refining,bottaro2020integrating,bergonzo2022conformational}.

Importantly, there are connections between the fields of RNA dynamics and RNA structure prediction.
Protein structure prediction has been recently revolutionized by the advent of AlphaFold \cite{jumper2021highly},
which has also been suggested to be useful in dissecting protein dynamics \cite{fowler2022accuracy}.
The use of deep-learning methods in RNA structure prediction has been limited to date \cite{townshend2021geometric}.
All the techniques developed for integrating MD simulations and solution experiments might be transferred to
such tools when/if they will be widely available and tested.

In conclusion, the combination of MD simulations and experimental data can be seen from a dual perspective.
From the computational side, experiments increase the quality
of the generated dynamical ensembles. From the experimental side, simulations are
a tool for interpreting the averaged signals obtained from solution measurements.
Overall, MD simulations and experimental information
can be synergistically combined filling their respective gaps, giving access to
detailed and accurate descriptions of RNA dynamics that are instead unreachable using a single approach.

\section*{Acknowledgements}

Sandro Bottaro and Serdal Kirmizialtin are acknowledged for reading the manuscript and providing useful suggestions.

\bibliographystyle{elsarticle-num}
\bibliography{main}

\end{document}